\documentclass[noshowpacs,amsmath,
twocolumn,
superscriptaddress,
8pt,
aps,prb]{revtex4-2}
\usepackage{setspace}
\usepackage{amsmath}
\usepackage{multirow}
\usepackage{graphicx}

\renewcommand{\figurename}{\textbf{Fig.}}

\usepackage{verbatim}
\usepackage{amsfonts}
\usepackage{amssymb}
\usepackage{epstopdf}
\usepackage{dcolumn}
\usepackage{xcolor}
\usepackage{verbatim}
\usepackage{hyperref}
\usepackage{siunitx}
\usepackage[version=4]{mhchem}
\usepackage{textcomp}
\setcitestyle{super}
\DeclareGraphicsExtensions{.pdf,.eps,.png,.jpg,.mps}
\begin{document}
\title{Microresonator-referenced soliton microcombs with zeptosecond-level timing noise}

\author{Xing Jin$^{1*}$, Zhenyu Xie$^{1*}$, Xiangpeng Zhang$^{2*}$, Hanfei Hou$^{1*}$, Fangxing Zhang$^{3}$, Xuanyi Zhang$^{1}$, Lin Chang$^{2}$, Qihuang Gong$^{1,3,4}$, and Qi-Fan Yang$^{1,3,4\dagger}$\\
$^1$State Key Laboratory for Artificial Microstructure and Mesoscopic Physics and Frontiers Science Center for Nano-optoelectronics, School of Physics, Peking University, Beijing 100871, China\\
$^2$State Key Laboratory of Advanced Optical Communications System and Networks, School of Electronics, Peking University, 100871 Beijing, China\\
$^3$Peking University Yangtze Delta Institute of Optoelectronics, Nantong, Jiangsu 226010, China\\
$^4$Collaborative Innovation Center of Extreme Optics, Shanxi University, 030006, Taiyuan, China\\
$^{*}$These authors contributed equally to this work.\\
$^{\dagger}$Corresponding author: leonardoyoung@pku.edu.cn}

\maketitle
{\bf \noindent Optical frequency division relies on optical frequency combs to coherently translate ultra-stable optical frequency references to the microwave domain \cite{fortier2011generation,li2014electro,xie2017photonic,yao2020optical,nakamura2020coherent}. This technology has enabled microwave synthesis with ultralow timing noise, but the required instruments are too bulky for real-world applications. Here, we develop a compact optical frequency division system using microresonator-based frequency references \cite{matsko2007whispering,alnis2011thermal,liu202236} and comb generators \cite{Kippenberg2018}. The soliton microcomb formed in an integrated \ce{Si3N4} microresonator is stabilized to two lasers referenced to an ultrahigh-$Q$ \ce{MgF2} microresonator. Photodetection of the soliton pulse train produces 25 GHz microwaves with absolute phase noise of -141 dBc/Hz (547 zs Hz$^{-1/2}$) at 10 kHz offset frequency. The synthesized microwaves are tested as local oscillators in jammed communication channels, resulting in improved fidelity compared with those derived from electronic oscillators. Our work demonstrates unprecedented coherence in miniature microwave oscillators, providing key building blocks for next-generation timekeeping, navigation, and satellite communication systems \cite{betz2015engineering,kodheli2020satellite}.}

\medskip
\noindent{\bf Introduction}

\noindent Consumer microwave technologies are transitioning towards X-band and beyond to accommodate the booming data traffic. However, these expanded radiofrequency channels will soon be overwhelmed by the exponential growth of satellite communications \cite{hecht2016bandwidth}. Furthermore, other microwave applications, including geodesy \cite{schluter2007international}, meteorological observations \cite{doviak2006doppler}, and marine monitoring \cite{huang2017ocean} are utilizing these same frequency ranges. It is envisioned that the improvement of channel capacity and transmission fidelity would be the most significant task for the widespread implementation of next-generation communication technologies.

To this end, low-noise microwave oscillators have become the cornerstone for advanced communication systems, which can enable high modulation formats and minimize inter-channel cross-talks. They also play a critical role in enhancing the accuracy, sensitivity, and dynamic range of radars \cite{ayhan2016impact}. While electronic oscillators are facing bandwidth limitations and degraded performance at higher-frequency bands, photonic technologies can overcome this challenge by synthesizing microwaves through laser beatnotes, as in the form of optoelectronic oscillators \cite{maleki2011optoelectronic,tang2018integrated}, Brillouin oscillators \cite{li2013microwave,gundavarapu2019sub}, dual-frequency lasers \cite{kittlaus2021low}, and optical frequency combs \cite{diddams2020optical}. Notably, optical frequency combs offer the unique ability to coherently connect optical and microwave signals through optical frequency division (OFD) \cite{yao2020optical}. This is usually achieved by stabilizing a set of equidistant comb lines -- the comb structure -- to optical frequency references, and the microwave-rate beatnote of comb lines sustains the relative stability of frequency references with remarkable fidelity \cite{fortier2011generation,li2014electro,xie2017photonic,nakamura2020coherent,li2023small,tetsumoto2021optically}. OFD of ultra-stable frequency references derived from atoms, ions, or high-finesse optical resonators has led to many performance milestones in microwave technology, including record low absolute timing noise of 41 zs Hz$^{-1/2}$ \cite{xie2017photonic} and fractional stability of 10$^{-18}$ \cite{nakamura2020coherent}.

\begin{figure*}
    \centering
    \includegraphics[width=\linewidth]{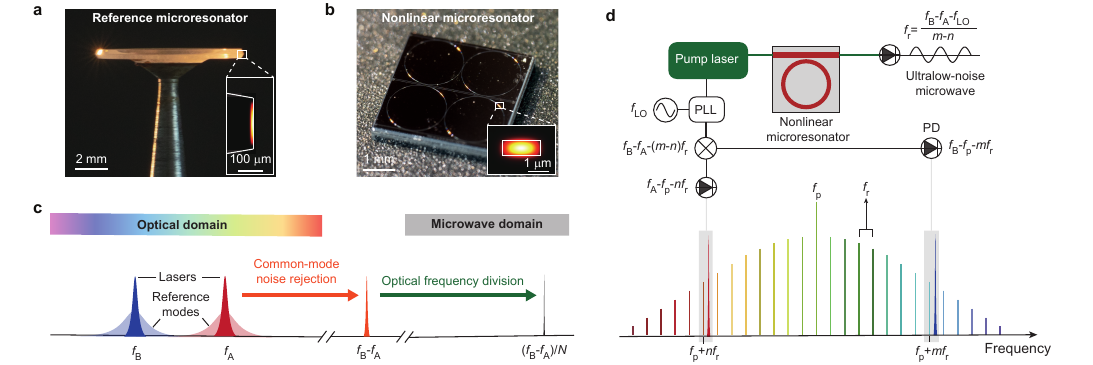}
    \caption{{\bf Conceptual illustration of microresonator-based optical frequency division (OFD).} {\bf a,} Photograph of the \ce{MgF2} reference microresonator. {\bf b,} Photograph of the \ce{Si3N4} nonlinear microresonator. The insets show the cross-sectional profile of the fundamental mode of each microresonator. {\bf c,} Concept of two-point OFD. {\bf d,} OFD error signal generation and feedback control of the soliton microcomb. PLL: phase-locked loop. PD: photodetector.}
    \label{Fig1}
\end{figure*}

The key components of an OFD system are frequency references and optical frequency combs, which are typically bulky and expensive. For example, conventional ultra-stable resonators, such as Fabry-Pérot resonators made of ultra-low-expansion glass or single-crystal silicon \cite{diddams2001optical,fortier2011generation,xie2017photonic,kessler2012sub}, often require vacuum enclosures or cryogenic temperatures to minimize acoustic noise and thermal drifts. Moreover, table-top optical frequency combs that are constructed from free-space or fiber-optic elements are usually too delicate for operation outside laboratories. Recently, integrated OFD systems have been proposed \cite{kudelin2023photonic,sun2023integrated}, which are motivated by the demonstration of soliton microcombs in microresonators \cite{Kippenberg2018}. Ultra-stable resonators have also been redesigned to operate under ambient conditions. Miniaturized OFD systems using centimeter-scale mini Fabry-Pérot resonators have achieved remarkable timing noise performance approaching 1 as Hz$^{-1/2}$ at 10 kHz offset frequencies \cite{kudelin2023photonic}.

Whispering-gallery-mode (WGM) microresonators have been proposed as potential candidates for ultra-stable frequency references. Firstly, these microresonators are solid-state devices with exceptional mechanical stability. The optical modes are confined along the periphery through total internal reflections, resulting in a highly compact geometry that can be realized using thin disks or photonic chips. This mechanism also supports ultra-high-$Q$ factors exceeding 100 million \cite{jin2021hertz,lee2012chemically,wang2013mid,liang2015high,yao2022soliton,puckett2021422}. Moreover, they can be fabricated using temperature-insensitive materials (with low thermo-optic and thermal expansion coefficients). These features provide appealing noise performance and compactness for OFD systems.

In this work, we advance miniature OFD technology by referencing a soliton microcomb to a WGM microresonator, and the synthesized microwave features record-low timing noise among all miniature photonic microwave oscillators. As depicted in Fig. \ref{Fig1}a, the reference microresonator is fabricated from a \ce{MgF2} disk through meticulous mechanical grinding and polishing processes \cite{qu2023fabrication}, yielding a diameter of 9.6 mm and a free-spectral-range ($FSR$) of 7 GHz. The trapezoid cross section supports WGMs with mode areas over 1000 \textmu m$^2$. Such a large mode volume, combined with a low thermo-optic coefficient of $1\times 10^{-6}$ K$^{-1}$, provides a low thermorefractive noise that is at least 30 dB lower than silica and \ce{Si3N4} microresonators with the same $FSR$. On the other hand, \ce{Si3N4} microresonators are very suitable for soliton microcomb generation due to their strong Kerr nonlinearity and tight optical confinement. The \ce{Si3N4} microresonator used here is fabricated via standard photolithography and plasma-etching processes \cite{ye2023foundry}, resulting in an $FSR$ closely approximating 25 GHz. Two-point OFD operations based on these two devices are described in Fig. \ref{Fig1}c. Two lasers ($f_A$ and $f_B$) are simultaneously stabilized to two modes of the \ce{MgF2} microresonator. The spatial overlap between these two modes provides common-mode noise rejection once the residual locking noise is below the thermal noise. It results in a mutual phase noise for $f_B-f_A$ that is lower than that of each laser, and we use $f_B-f_A$ as the frequency reference for further division. As described in Fig. \ref{Fig1}d, the microresonator is excited by a continuous-wave pump laser to generate a soliton microcomb with repetition frequency $f_r$. The reference lasers are combined with two adjacent comb lines that are indexed by $m$ and $n$ relative to the pump. The photodetected beatnotes are then down-mixed to $f_B-f_A-(m-n)f_r$ and locked to a local oscillator (LO) with the frequency of $f_\mathrm{LO}$. This locking is commonly achieved by adjusting the frequency of the pump laser, which regulates the repetition frequency through the Raman-induced soliton-self-frequency-shift mechanism \cite{yi2016theory}. Photodetection of the soliton pulse train provides the stabilized repetition frequency as:
\begin{equation}
    f_r=\frac{f_B-f_A-f_\mathrm{LO}}{m-n}.
\end{equation}
If the phase noise of the LO is not dominant, we can anticipate that the phase noise of $f_r$ is reduced by a factor of $(m - n)^2$ compared with that of $f_B - f_A$.

\medskip
\noindent{\bf Results}

\begin{figure*}
\centering
\includegraphics[width=\linewidth]{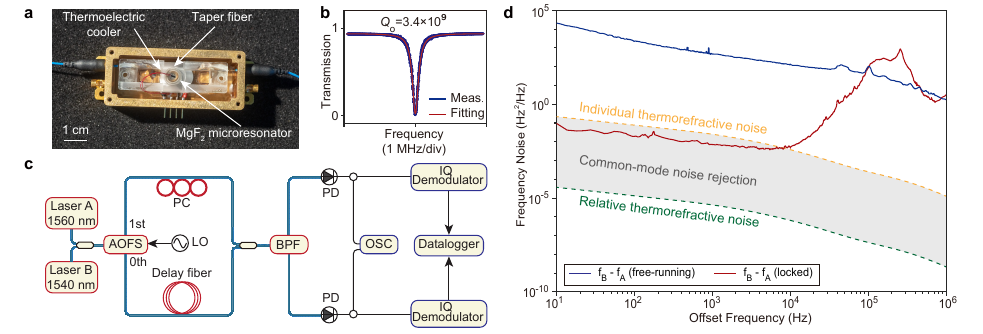}
\caption{{\bf Characterization of optical frequency references.} {\bf a, } Photograph of the packaged \ce{MgF2} reference microresonator. {\bf b,} Typical transmission spectrum of a resonance. The intrinsic quality factor is derived from Lorentzian fitting as 3.4 billion. {\bf c,} Setup of multi-frequency delayed self-heterodyne interferometer. AOFS: acousto-optic frequency shifters; PC: polarization controller; LO: local oscillator; BPF: optical bandpass filter;  OSC: oscilloscope. {\bf d,} Measured relative frequency noise of the free-running and locked reference lasers. The simulated individual and relative thermorefractive noise of the references are also plotted, between which is the regime accessible via common-mode noise rejection.}
\label{Fig2}
\end{figure*}

\noindent{\bf Optical frequency references}

\noindent We first characterize the performance of the frequency reference. To ensure stable operation, we enclose the \ce{MgF2} microresonator in a metallic package with a tapered fiber coupler and a thermoelectric cooler (Fig. \ref{Fig2}a). The packaging process does not degrade the resonator's $Q$ factors, which can exceed 3 billion under near-critical coupling conditions. Two fiber lasers at wavelengths of 1540 nm and 1560 nm are stabilized to the \ce{MgF2} microresonator using the Pound-Drever-Hall (PDH) technique. Servo control is applied to the laser piezos and acousto-optic frequency shifters (AOFS), providing a locking bandwidth of up to 300 kHz. The relative frequency noises of these lasers are characterized using a multi-frequency delayed self-heterodyne interferometer \cite{jeong2020ultralow,lao2023quantum} (Fig. \ref{Fig2}c). The two reference lasers are combined and sent to an AOFS driven by a 55 MHz LO. The 0th and 1st-order diffraction lights travel in two arms of an unbalanced Mach-Zender interferometer respectively before again being combined and separated by a bandpass filter according to their wavelength. The two reference lasers, along with their frequency-shifted counterparts, are detected by two photodetectors, yielding two 55 MHz beatnotes containing the phase fluctuation of the respective reference lasers. Simultaneous extraction of the phases of the beatnotes is realized through two methods. Hilbert transform of the oscilloscope traces provides fast sampling rates but limited recording length; acquisition of the IQ demodulated beatnotes using a datalogger provides a very long recording length at limited sampling rates ($\textless$800 kHz). Such a combination allows for noise measurement bandwidth from 10 Hz to 1 MHz.  The relative frequency noise of the reference lasers is shown in Fig. \ref{Fig2}d, where locking to the microresonator results in a noise reduction of more than 50 dB. Notably, the relative noise at offset frequencies from 10 Hz to 10 kHz is 10 dB below the thermorefractive noise of the microresonator. This is attributed to common-mode noise rejection, which can lead to an additional 30 dB reduction factor if residual locking noises are further minimized.

\begin{figure*}
\centering
\includegraphics[width=\linewidth]{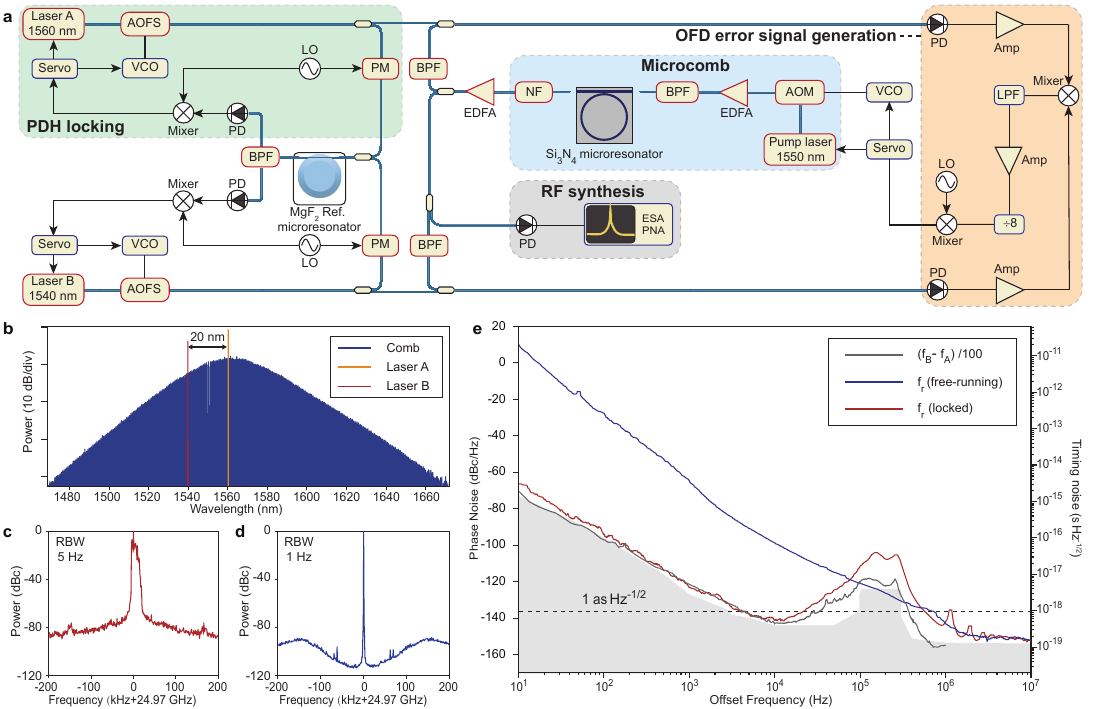}
\caption{{\bf Optical frequency division characterization.} {\bf a,} Experimental setup for OFD. VCO: voltage-controlled oscillator; PM: phase modulator; EDFA: erbium-doped fiber amplifier; NF: notch filter; Amp: electrical amplifier; LPF: electrical lowpass filter; ESA: electrical spectrum analyzer; PNA: phase noise analyzer. {\bf b,} Optical spectra of the soliton microcomb and the reference lasers. {\bf c,} Beatnote of the free-running soliton microcomb. {\bf d,} Beatnote of the locked soliton microcomb. RBW: resolution bandwidth. {\bf e,} Single sideband phase noise of the free-running and locked beatnotes of the soliton microcomb. The projected contribution (scaled by 40 dB) from the optical frequency references is also shown for comparison. The dashed black line indicates the timing jitter level of 1 as Hz$^{-1/2}$. The gray shaded area represents the noise floor of the PNA during the measurement.}
\label{Fig3}
\end{figure*}

\medskip
\noindent{\bf Optical frequency division}

\noindent The detailed experimental setup is illustrated in Fig. \ref{Fig3}a. Soliton microcombs are generated by pumping a packaged \ce{Si3N4} microresonator at 1550 nm using an amplified fiber laser. By manually tuning the frequency of the pump laser from a multi-soliton state \cite{guo2017universal}, a single soliton state is achieved, exhibiting a characteristic spectral envelope of sech$^2$ and a 3-dB bandwidth spanning over 30 nm (Fig. 3b). This bandwidth is sufficient to bridge the frequency gap between reference lasers (20 nm, or equivalently 2.5 THz), and dividing it to the 25 GHz repetition frequency should provide a noise reduction factor of 40 dB. We use bandpass filters to select the frequency components near 1540 nm and 1560 nm of the amplified soliton microcomb to beat with the reference lasers separately. The beatnotes are amplified and then down-mixed to generate the OFD error signal. To increase the locking range, the OFD error signal is electrically divided by a factor of 8 before sending it into the PLL. By adding an AOFS to the pump laser, the locking bandwidth can be extended to 200 kHz. Finally, the amplified soliton microcomb is sent into a fast photodetector with proper dispersion compensation to suppress the Gordon-Haus jitter introduced by the amplification process \cite{gordon1986random}.

The radiofrequency beatnotes of the free-running and stabilized soliton microcombs acquired with an electrical spectral analyzer are compared in Fig. \ref{Fig3}c and d. The former is susceptible to various technical noise sources, including noise transduced from the pump laser and temperature fluctuations within the microresonator, leading to a broader linewidth and frequency drift. In contrast, the repetition frequency of the stabilized soliton microcomb primarily follows the frequency references and exhibits a significantly narrowed linewidth. Figure \ref{Fig3}e presents a more quantitative comparison conducted using a phase noise analyzer (Rohde \& Schwarz FSWP50). Within an offset frequency range of 10 Hz to 10 kHz, OFD leads to a noise reduction exceeding 40 dB compared to the free-running case. Within this frequency range, the phase noise of the stabilized repetition frequency aligns closely with that of the divided frequency references, achieving a level of -123(-141) dBc/Hz at 1(10) kHz offset frequency. Above the corner frequencies of approximately 200 kHz, its phase noise converges to the shot-noise-limited level of -152 dBc/Hz. 

\begin{figure*}
\centering
\includegraphics[width=\linewidth]{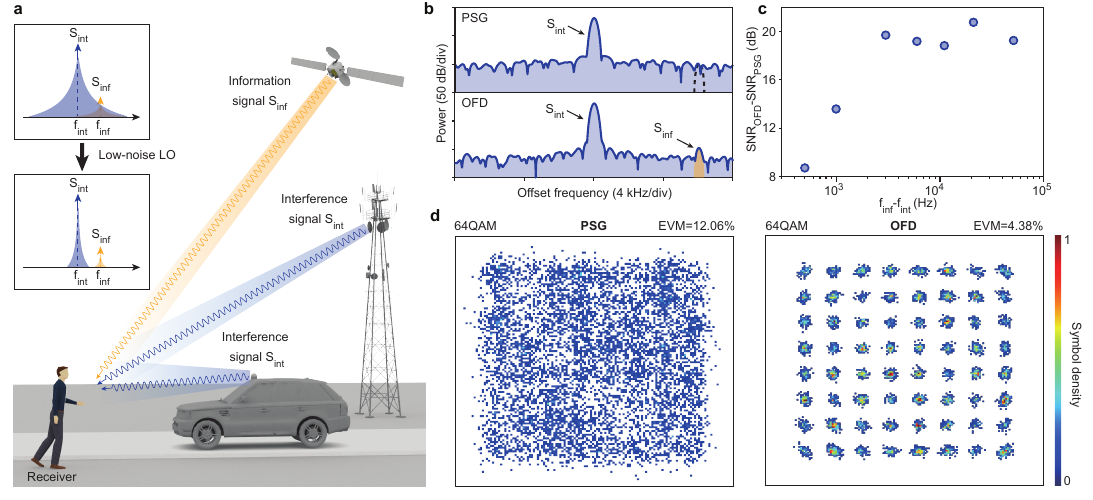}
\caption{{\bf Anti-interference experiments.} {\bf a,} Conceptual diagram of an interfered communication channel. {\bf b,} Electrical spectra of the signals after up-conversion using Keysight PSG E8257D (PSG) and OFD-based LOs. The black dashed line in the up panel indicates the inferred information signal. {\bf c,} The differences in signal-to-noise ratio (SNR) of the information signal for PSG and OFD-based LOs. The x-axis is the frequency gap between the information and interference signals. {\bf d,} Constellation diagram of interfered 64QAM data transmission experiments using PSG and OFD-based LOs. The normalized symbol density is indicated by color.}
\label{Fig4}
\end{figure*}

\medskip
\noindent{\bf Anti-interference experiments}

\noindent We compare the performance of our OFD system with a high-end electronic oscillator (Keysight PSG E8257D, referred to as PSG in the following text) as the LOs in two anti-interference experiments. The objective is to test the anti-interference performance of the LOs under conditions of intense communication jamming \cite{CARR2000163}, such as satellite communications disturbed by millimeter waves for 5G communications and autonomous vehicles (Fig. \ref{Fig4}a). We start the experiments by up mixing a pair of frequency-close single-tone signals -- one serving as the strong interference (-20 dBm) and the other as the weak information (-97 dBm) -- with the local oscillation signal generated by PSG and OFD respectively (Extended Data Fig. \ref{ExtenFig1} a). As shown in Fig. \ref{Fig4}b, when PSG is used as the LO, the noise floor of the converted interference is so high that the information signal is drowned out. When shifting from PSG to OFD-based LO, a previously obscured information signal can be observed, which can be attributed to the ultra-low phase noise characteristics of the OFD-based LO. We then compare the signal-to-noise ratio of the information signals for the two cases at different frequency separations. Notably, our OFD system provides a 20 dB advantage over the PSG when the two signals are 10 kHz apart (Fig. \ref{Fig4}c).

We then replace the single-tone information signal with communication data in 64QAM format with a symbol rate of 50 kBd to imitate a real communication scenario (Extended Data Fig. \ref{ExtenFig1} b). The frequency gap between the information and interference signals is set to be 35 kHz. The up-mixed information signal is analyzed using a vector signal analyzer for constellation diagram construction and error vector magnitude (EVM) measurement. As can be seen in Fig \ref{Fig4}d, using the PSG-based LO yields an EVM of 12.06\%, exceeding the 8\% threshold for standard 64QAM telecommunication. In contrast, using the OFD-based LO yields an EVM of only 4.38\%, maintaining an adequate signal fidelity under heavy jamming due to its lower phase noise. These experimental results suggest that our system can potentially outperform table-top electronic microwave oscillators in the quest for robust communication systems.

\medskip

\noindent{{\bf Conclusion and outlook}}

\begin{figure}
\centering
\includegraphics[width=\linewidth]{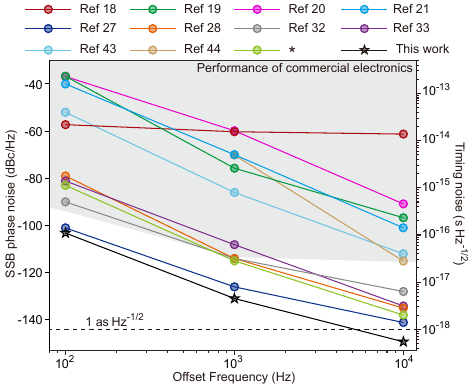}
\caption{{\bf Comparison of microwaves synthesized by miniature photonic microwave oscillators, scaled to 10 GHz.} Performance of an integrated optoelectronic oscillator \cite{tang2018integrated}, OEwaves HI-Q X-band optoelectronic oscillator (*), a Brillouin laser oscillator in silica \cite{li2013microwave}, a Brillouin laser oscillator in \ce{Si3N4} \cite{gundavarapu2019sub}, a self-injection-locked laser based oscillator \cite{kittlaus2021low}, a free-running microcomb in \ce{Si3N4} \cite{liu2020photonic}, a free-running microcomb in \ce{MgF2} \cite{liang2015high}, a free-running microcomb in silica \cite{yao2022soliton}, an optical parametric oscillation in \ce{Si3N4} referenced microcomb in \ce{Si3N4} \cite{zhao2023all}, a \ce{Si3N4} coil cavity referenced microcomb in \ce{Si3N4} \cite{sun2023integrated}, a mini Fabry–Pérot cavity referenced microcomb in \ce{Si3N4} \cite{kudelin2023photonic}, and a \ce{MgF2} microresonator referenced microcomb in \ce{Si3N4} (this work) are in comparison. All of the noise performance in this figure is scaled to a 10 GHz carrier frequency. The gray shaded area indicates the noise performance of commercial electronics (Keysight PSG E8257D). The dashed black indicates the timing jitter level of 1 as Hz$^{-1/2}$.}
\label{Fig5}
\end{figure}

\noindent The phase noise (scaled to 10 GHz carrier) and absolute timing noise of miniature photonic microwave oscillators are summarized in Table \ref{Table1} and Fig. \ref{Fig5}. The benefit of introducing frequency references is very obvious: it reduces the absolute timing noise of free-running oscillators by more than 10 dB to a level that is not attainable by commercial electronic oscillators. Notably, our findings represent the lowest timing noise achieved from 100 Hz to 10 kHz offset frequencies among all miniature photonic microwave oscillators. A significant milestone is the realization of timing noise below 546 zs Hz$^{-1/2}$ at a 10 kHz offset frequency. Such zeptosecond timing noise regime was previously only accessible by referencing microcombs to a table-top ultra-stable laser \cite{lucas2020ultralow}.

While the state-of-the-art table-top OFD systems provide more than an order of magnitude lower timing noise \cite{xie2017photonic}, such performance difference can be mitigated by increasing the division ratio, which would reduce the impact of the residual locking noise. One-point OFD can be anticipated using self-referenced microwave-rate microcombs, which could be achieved through a combination of dispersion engineering \cite{anderson2022zero} and efficient pumping schemes \cite{helgason2023surpassing}. Additionally, optimizing the residual amplitude modulation in the phase modulator may improve the level of common-mode noise rejection, potentially providing over 20 dB lower timing noise as inferred from Fig. \ref{Fig2}d.

Further integration of the OFD system is viable. By leveraging thin-film lithium niobate technology \cite{boes2023lithium}, key components such as phase modulators \cite{wang2018integrated} and acousto-optic modulators \cite{shao2020integrated} can be implemented on chips. Integrated lasers \cite{gundavarapu2019sub,jin2021hertz,li2021reaching} and soliton microcombs \cite{stern2018battery,raja2019electrically,shen2020integrated,jin2021hertz,xiang2021laser} with competitive noise performance have also been recently demonstrated, and their low-noise amplification for photodetection can be realized using erbium-doped waveguide amplifiers \cite{liu2022photonic}. In addition, electro-optic comb generators can also be chosen as the frequency divider such that the desired microwaves can be directly read out from the driving VCO \cite{li2014electro,li2023small,wang2018integrated,hu2022high}. As for the optical references, integrated waveguide couplers have been developed to interface with ultra-high-$Q$ crystalline microresonators \cite{liu2018low,anderson2018highly}. They should enable laser self-injection locking to the microresonators, such that the PDH locking setup can be eliminated to reduce the form factors and electrical power consumption \cite{liang2015ultralow}. These advances would extend the impact of OFD technologies to consumer markets such as high-resolution radars for autonomous vehicles and transceivers for high-speed wireless communications, providing a combination of exceptional performance, compact size, robust operation, and mass production. They can also facilitate a number of scientific tasks, including very long baseline interferometry for space-based radio astronomy \cite{akiyama2022first} and high-fidelity manipulation of superconducting qubits \cite{clarke2008superconducting,gong2021quantum}. 

\begin{table*}
\includegraphics[width=\linewidth]{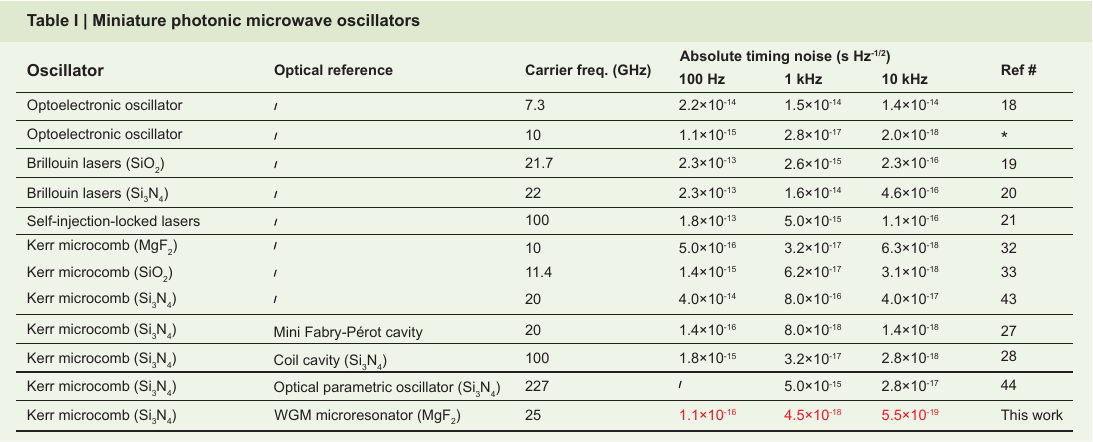}
\caption{{\bf Comparison of miniature photonic microwave oscillators. } * OEwaves HI-Q X-band optoelectronic oscillator. The material to form the resonators that serve as the oscillators and optical references are indicated in the table.}
\label{Table1}
\end{table*}

\begin{footnotesize}
\end{footnotesize}

\medskip
\noindent\textbf{Methods}

\begin{footnotesize}
\noindent{\bf Simulation of thermorefractive noise.} The thermorefractive noise is numerically simulated using a finite-element method based on fluctuation-dissipation theorem \cite{kondratiev2018thermorefractive,yang2021dispersive}. The parameters of \ce{MgF2} used in the simulations are the heat capacity $920$ $\SI{}{\J\ \kg^{-1}\ \K^{-1}}$, thermal conductivity $20.98$ $\SI{}{\W\ \m^{-1}\ \K^{-1}}$, material density $3.18\times10^3$ $\SI{}{\kg \ \m^{-3}}$, refractive index 1.37, and thermal-optic coefficient $1\times 10^{-6}$ $\SI{}{\K^{-1}}$. In the simulations, the ambient temperature is assumed to be $300$ $\SI{}{\K}$. 

\end{footnotesize}
\medskip
% \noindent{\bf Communication experiments.}

% \noindent\textbf{Data availability}
% \begin{footnotesize}
% \noindent The data that support the plot within this paper and other findings of this study are available at . 
% \end{footnotesize}

% \medskip

% \noindent\textbf{Code availability}
% \begin{footnotesize}
% \noindent The codes that support the findings of this study are available at . 
% \end{footnotesize}

\medskip
\noindent\textbf{Acknowledgments}

\begin{footnotesize}
\noindent The \ce{Si3N4} chips used in this work were fabricated by Qaleido Photonics. The \ce{MgF2} microresonators used in this work were fabricated and packaged by Peking University Yangtze Delta Institute of Optoelectronics. The project is supported by National Key R\&D Plan of China (Grant No. 2021YFB2800601), Beijing Natural Science Foundation (Z210004), National Natural Science Foundation of China (92150108,62305006), Nantong Science and Technology Bureau (MS12022003), and the High-performance Computing Platform of Peking University.
\end{footnotesize}
\medskip

\noindent\textbf{Author contributions} 

\begin{footnotesize}
\noindent Experiments were conceived and designed by X.J., Z.X., X.Z., H.H. and Q-F.Y. Measurements and data analysis were performed by X.J., Z.X., X.Z., H.H., and Q-F.Y. Numerical simulations of thermorefractive noise were performed by X.J., H.H, and X.Z. The \ce{MgF2} microresonator was designed and fabricated by X.Z. and F.Z. All authors participated in preparing the manuscript.
\end{footnotesize}

\medskip

\noindent\textbf{Competing interests}

\begin{footnotesize}
\noindent The authors declare no competing interests.
\end{footnotesize}

\medskip

\noindent\textbf{Additional information}

\begin{footnotesize}
\noindent Correspondence and requests for materials should be addressed to Q-F.Y.
\end{footnotesize}

\renewcommand{\figurename}{\textbf{Extended Data Fig.}}

\begin{figure*}
\centering
\setcounter{figure}{0}
\includegraphics[width=\linewidth]{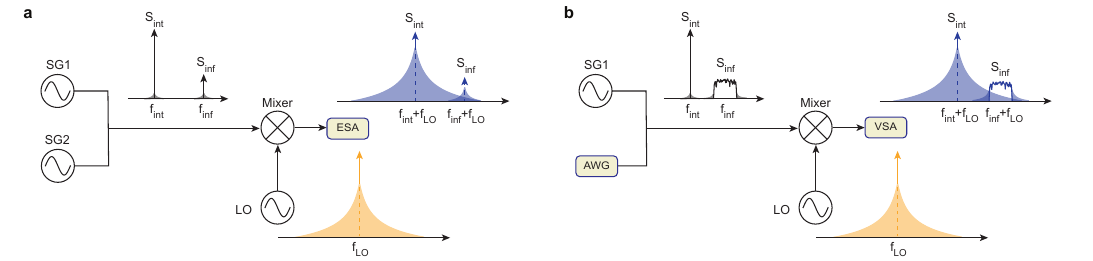}
\caption{{\bf Experiments setup for anti-interference experiments.} {\bf a,} The two signal generators produce a strong interference signal and a weak information signal respectively. The two signals are coupled together, mixed with a LO, and then sent to an electrical signal analyzer (ESA). {\bf b,} A signal generator and an arbitrary waveform generator (AWG) produce a strong interference signal and a baud information signal for data transmission respectively. The two signals are coupled together, mixed with a LO, and then sent to a vector signal analyzer (VSA) to obtain the constellation diagrams.}
\label{ExtenFig1}
\end{figure*}

\bibliography{ref}
\end{document}